\documentclass[twocolumn, showpacs, amsmath, amssymb, prb]{revtex4}

\usepackage{amssymb}
\usepackage{amsmath}
\usepackage{amsfonts}
\usepackage{graphicx}

\begin{document}

\title{Acceleration of Metal Nanoparticle with Irradiation Pressure}

\author{Nicolas I.~Grigorchuk \footnote{email: ngrigor@bitp.kiev.ua}}
\affiliation{Bogolyubov Institute for
Theoretical Physics, National Academy of Sciences of Ukraine, \\
14-b Metrologichna Str., Kyiv-143, Ukraine, 03143}


\begin{abstract}
The acceleration of an spheroidal metal nanoparticle in an
irradiation field with a frequency close to the surface plasmon
vibration has been considered. Under the action of radiation pressure,
the polarizability for nonspherical particle becomes a tensor quantity.
The analytical expressions for the resonance acceleration components
for the cases of plane-polarized and circularly polarized light have
been derived. We have demonstrated that the resonance acceleration
can depend substantially on the shape of a metal nanoparticle
and its orientation with respect to the directions of light
propagation and the light polarization.
\end{abstract}

\maketitle

\section{Introduction}
The advent of lasers made the development of researches in the
field of microparticle trapping, confinement, and manipulation
possible\cite{II}. The hot-electron pressure triggers the
anisotropic shape oscillations due to the thermal
expansion of the optically heated particles\cite{PGM}.
The resonant radiation pressure on neutral particles was
investigated in\cite{GSG}. Rawson and May observed
the angular stabilization of matter by radiation\cite{RM}
which is familiar to trapping of small particles by
radiation pressure. In 1970, Arthur Ashkin\cite{AA}
has been demonstrated, for the first time, the trapping
and the manipulating of a micron-sized dielectric spherical
particle in the field of two opposing laser beams.
An another work\cite{AD} was devoted to the observation of
resonances in the radiation pressure on dielectric spheres.
The plasmon-resonance conditions for optical forces on small
particles was than considered in detail by Arias-Gonz{\'a}lez
and Nieto-Vesperinas\cite{AN}. More latter it was developed
the theory of optical tweezers\cite{NBX}.
Much efforts to develop the photonic force spectroscopy on
metal nanoparticles (MNs) was applied by P.~Chaument with
coauthors\cite{CRN}.

For recent years, intensively developed have been both the
researches of the peculiarities inherent to the mechanisms of
light pressure action upon nanoparticles and the implication
of this action in the tasks of small particle manipulation.
Such applications meet a wide usage in biology, medicine,
and microelectronics.

A theory for the generation in a spheroidal MN of an angular
momentum under the action of ultrashort laser pulse is
developed in the work\cite{GJ}. The optical radiation force
on a dielectric sphere illuminated by a linearly polarized
Airy light-sheet was studied recently in\cite{SLS}. A review
of some relevant problems can be found, e.g., in the work\cite{SD}.

A theoretical study of the time-averaged force exerting upon a
spherical particle in a time-harmonic-varying electromagnetic
field has been carried out in the work\cite{AN}. The expression
obtained there for the force components depends on the gradient
of the electromagnetic wave intensity and on the particle's
polarizability. The particle was considered spherical, so that
its polarizability was characterized by a scalar parameter.

In present work, we consider MNs of the spheroidal form.
In this case, the particle polarizability becomes a tensor
and can depend rather strongly on the particle's
morphology \cite{TG}. Moreover, the high-frequency (optical)
conductivity\cite{G1}, which is connected to the imaginary
part of particle's polarizability and defines its absorption,
also becomes a tensor. The dependence of the polarizability
of MN on its form becomes especially appreciable in the
infra-red range of frequencies. As a result,
the expression for the components of the MN acceleration
with action of the laser beam, would differ substantially
from those obtained in the spherical case.

\section{Formulation of the Problem}

For particles, whose dimensions are considerably smaller than the length
of the electromagnetic wave, we apply the Rayleigh approximation, i.e.
the particle is considered as a dipole in a non-uniform field.
The force affecting such a particle equals
\begin{equation}
 {\bf{F}}=({\bf{P}}\cdot {\bf{\nabla}}){\,\bf{E}}+{\frac{1}{c}}
  \dot{\bf{P}} \times {\bf{B}},
   \label{eq1}
\end{equation}
where $\bf{P}$ is the dipole moment of the particle,
$\bf{E}$ the electric and $\bf{B}$ the magnetic fields,
and $c$ the speed of light. All quantities in Eq.~(\ref{eq1})
are real. For our purpose, it is convenient to use
the complex ones, using in Eq.~(\ref{eq1}) for an
arbitrary vector $(\bf V)$ the conditional scheme
\begin{equation}
 {\bf V}\Rightarrow {\frac{1}{2}} ( {\bf V} + {\bf V}^{\ast} ).
  \label{eq2}
\end{equation}
We will suppose that the complex quantities
are depended on time harmonically:
\begin{equation}
 {\bf V}^{\ast} = {\bf{V}}_0 \exp(- i\omega t).
  \label{eq3}
\end{equation}
Here $\omega$ is the frequency of the electromagnetic wave.
Now, we can introduce the electromagnetic acceleration
averaged over the time period~$T$:
\begin{eqnarray}
 \overline{\bf{\cal A}} &=&
  \frac{1}{4TM}\int\limits_{-T/2}^{T/2} {dt}
   \Bigl\{[({\bf{P}} + {\bf{P}}\,^{\ast})\cdot {\bf{\nabla}}]\,
    ({\bf{E}} + {\bf{E}}^{\ast} )  \Bigr. \nonumber \\&\times&\Bigl.
     \frac{1}{c}(\dot {\bf{P}} +
      \dot {\bf{P}}^{\ast})\times ({\bf{B}} + {\bf{B}}^{\ast}) \Bigr\},
       \label{eq4}
\end{eqnarray}
where $M$ is the mass of the particle.
The second term in the integrand of expression (\ref{eq4})
can be integrated by parts with using well-known ratio
\begin{equation}
 -{\frac{1}{c}}{\frac{d{\bf{B}}}{dt}} = {\rm rot}{\,\bf{E}}.
    \label{eq5}
     \end{equation}
Then, instead of Eq.~(\ref{eq4}), we obtain
\begin{eqnarray}
 \overline{\bf{\cal A}} &=&
  {\frac{1}{4MT}}{\int\limits_{-T/2}^{T/2} {dt}}\Bigl
   \{(({\bf{P}} + {\bf{P}} \,^{\ast}) \cdot {\bf{\nabla}} )({\bf{E}}
    + {\bf{E}}^{\ast} ) \Bigr.\nonumber
     \\& + &\Bigl.
      \frac{1}{c}({\bf{P}} + {\bf{P}}\,^{\ast} )
       \times[{\bf{\nabla}} \times ({\bf{E}} + {\bf{E}}\,^{ \ast }) ]
        \Bigr\}.
         \label{eq6}
\end{eqnarray}
Now, taking advantage of the explicit dependence on time
(see Eqs.~(\ref{eq3})), it is easy to carry out the integration
in Eq.~(\ref{eq6}) in time. The acceleration averaged
over wave period which is connected with the dipole
moment $P_0$ in the general case will have the form
\begin{eqnarray}
 \overline{\bf{\cal A}} &=& \frac{1}{4M}\Bigl\{ ({\bf{P}}_0
  \cdot {\bf{\nabla}}){\bf{E}}_0^{\ast} + ({\bf{P}}_0^{\ast}
   \cdot {\bf{\nabla}}){\bf{E}}_0
    \Bigr.\nonumber\\& + &\Bigl.
     {\bf{P}}_0 \times [{\bf{\nabla}
      \times} {\bf{E}}_0^{\ast} ] + {\bf{P}}_0^{\ast }
       \times [{\bf{\nabla}} \times {\bf{E}}_0] \Bigr\}.
        \label{eq7}
\end{eqnarray}
Later we will use formula (\ref{eq7}) to calculate the MN
acceleration under resonant irradiation by the laser beam.

Below, we consider a MN with an
ellipsoid-of-revolution form. In the reference frame connected
to the principal axes of this ellipsoid, the dipole moment
of such a particle looks like\cite{BH}
\begin{equation}
 P_{0j}={\frac{V}{4\pi}} { \frac{\varepsilon_{jj}-1}
 {1+L_j (\varepsilon_{jj}-1) }} E_{0j},
  \quad {j\ = \ x,\ y,\ z.}
   \label{eq8}
    \end{equation}
Here, $V$ is the volume of the MN,
$L_{j}$ are the depolarization factors,
\begin{equation}
  \varepsilon_{jj} = {\varepsilon}_{jj}^{\prime} +
   {\varepsilon}^{\prime \prime} = {\varepsilon}^{\prime} +
    i{\frac{4\pi}{\omega}}\sigma_{jj},
     \label{eq9}
      \end{equation}
$\varepsilon^{\prime}$ is the real part
of the dielectric constant which has the form
\begin{equation}
 {\varepsilon}^{\prime} = 1-{\frac{\omega_{pl}^{2}}{\omega^2}},
  \label{eq10}
   \end{equation}
$\omega_{pl}$ is the plasma oscillation frequency, and $\sigma_{jj}$ are
the diagonal elements of the tensor of high-frequency (optical) conductivity.

We admit the characteristic particle sizes are smaller than
the mean free path of an electron in the direction of its scattering by
phonons. Provided both such sizes and the asymmetric form of the MN,
the conductivity becomes a tensor value as was demonstrated in work\cite{G2}.
In turn, the conductivity and, therefore, dissipation are due to the
action of both the electric field $E$ (electric absorption) and the
magnetic field $B$ (magnetic absorption) of the electromagnetic wave.
In the case of MN with the form of an ellipsoid of revolution, the following
components of the tensor $\sigma_{jj}$ are distinct from zero in the
reference frame connected to the principal axes of this ellipsoid:
\begin{equation}
 \sigma_{xx} = \sigma_{yy}\equiv \sigma_{\bot },\quad
  \sigma_{zz}\equiv \sigma_{\parallel},
   \label{eq11}
    \end{equation}
while the depolarization factors equal
  \begin{equation}
   L_{x}(e_p) = L_{y}(e_p) = {\frac{1}{2}}[1-L_{z}(e_p)]\equiv L_{\bot},
   \label{eq12}
    \end{equation}

     \begin{equation}
      L_z(e_p)\equiv L_{||}=\left\{\begin{array}{ll}
      \frac{1-e^2_p}{2e^3_p}\left(\ln{\frac{1+e_p}{1-e_p}-2e_p}\right), &
       R_{\bot} < R_{\Vert} \\
        \frac{1+e^2_p}{e^3_p}(e_p-\arctan e_p), & R_{\bot} > R_{\Vert}
         \end{array} \right..
          \label{eq13}
           \end{equation}

In expressions (\ref{eq13}) the notation
 \begin{equation}
   e_p^2=\left\{
    \begin{array}{ll}
     1-R^2_{\bot}/R^2_{\Vert}, & R_{\bot} < R_{\Vert}, \\
      R^2_{\bot}/R^2_{\Vert}-1, & R_{\bot} > R_{\Vert},
       \end{array}
        \right.
         \label{eq14}
          \end{equation}
is used, where $R_{\parallel}$ and $R_{\perp}$ are the corresponding
semi-axes of the ellipsoid of revolution ($R_{\parallel}$
is along with revolution axis and $R_{\perp}$ is perpendicular to this axis).

Introducing the vector components of the dipole moment in the form
 \begin{equation}
  P_{0i} = {\sum\limits_{j}{\alpha_{ij}E_{0j}}},
   \label{eq15}
    \end{equation}
Eqs.~(\ref{eq8}) and (\ref{eq11}) yield the following expressions for
nonzero components of the polarizability tensor $\alpha_{jj}$:
\begin{equation}
 \alpha_{xx}=\alpha_{yy}\equiv\alpha_{\bot} = {\frac{V}{4\pi}}{
  \frac{(\varepsilon_{\bot}-1)}{1+L_{\bot}(\varepsilon_{\bot}-1)}},
   \label{eq16}
    \end{equation}
  \begin{equation}
   \alpha_{zz} \equiv \alpha_{\parallel}  =
    {\frac{V}{4\pi}}{\frac{(\varepsilon_{_{\parallel}} -1)}{{1 +
     L_{\parallel} (\varepsilon_{_{\parallel}} - 1)}}},
      \label{eq17}
       \end{equation}
where
\begin{equation}
 \varepsilon_{\parallel} = {\varepsilon}^{\prime}+i{
  \frac{4\pi}{\omega}}\sigma_{_{\parallel}},\quad
   \varepsilon_{\bot} = {\varepsilon}^{\prime} + i{\frac{4\pi}{\omega} }
    \sigma_{\bot} .
     \label{eq18}
      \end{equation}
The expressions for $\sigma_{\perp}$ and $\sigma_{\parallel}$ for
various specific conditions are presented in the work \cite{TG}.
In particular, if the electric absorption dominates, simple analytical
expressions for the components $\sigma_{\perp}$ and $\sigma_{\parallel}$
can be obtained in the cases of strongly prolate ($R_{\parallel}\gg
R_{\perp}$) and strongly oblate ($R_{\parallel}\ll R_{\perp}$)
ellipsoids \cite{TG}:
\begin{equation}
 \sigma_{\parallel}\approx {\frac{3}{2}}\sigma_{\bot}\approx{\frac{9\pi}
  {64}}{\frac{v_{\rm F}}{R_{\bot}}}{\frac{ne^2}{m\omega^{2}}}
   \quad (R_{\parallel}\gg R_{\bot}),
    \label{eq19}
     \end{equation}
\begin{equation}
 \sigma_{\parallel} \approx {\frac{1}{2}} \sigma_{\bot}
  \approx {\frac{9}{16}}{\frac{v_{\rm F}}{R_{\parallel}}}{
   \frac{ne^2}{m\omega^2}}
    \quad (R_{\parallel}\ll R_{\bot}) .
     \label{eq20}
      \end{equation}
Here, $v_{\rm F}$ is the Fermi velocity, $n$ the
concentration of electrons, and $m$ the electron mass.

For spherical MNs ($R_{\parallel} = R_{\perp} = R$),
we obtain
\begin{equation}
 \sigma_{\parallel} = \sigma_{\bot} =
  {\frac{3}{4}} { \frac{v_{\rm F}}{R} }
   {\frac{ne^2}{m\omega^2} }.
    \label{eq21}
     \end{equation}
Formulae (\ref{eq19}) and (\ref{eq20}) are valid in
the case of high-frequency fields, when the frequency
of light is higher than the transit-time ones
($\omega > v_{\rm F}/R_{\bot},v_{\rm F}/R_{\parallel}$).

Starting from formula (\ref{eq15}) with using Eq.~(\ref{eq11}),
the dipole moment can be written down for arbitrary
coordinate system in the form
\begin{equation}
 {\bf{P}}_0 = \alpha_{\bot}{\bf{E}}_0 + (\alpha_{\bot}-
  \alpha_{\parallel})({\bf{n}}
   {\bf{E}}_0){\bf{n}}.
    \label{eq22}
     \end{equation}
Here, $\bf{n}$ is a unit vector directed along the axis of
rotation of the ellipsoid. Formulae (\ref{eq22}) and (\ref{eq7})
will serve as the basic ones for studying the MN acceleration.

\section{Acceleration Under the Action of Light Pressure}

In order to obtain the explicit expression for the time-averaged acceleration
(\ref{eq7}), it is necessary to establish the coordinate dependence of
the field ${\bf{E}}_0$. As the first example of such a dependence, we
take this field in the form with a linear polarization along $x$-axis.
It looks like
\begin{equation}
 {\bf{E}}_{0} = (E_{x},0,0);  \quad
  E_{x} = E_{0} e^{-x^2/(2a^2)}e^{ikz},
   \label{eq23}
    \end{equation}
where $a$ is the radius of the light beam.
Substituting expressions (\ref{eq22}) and (\ref{eq23})
into Eqs.~(\ref{eq7}), we obtain the expressions for nonzero
components of the time-averaged MN acceleration:
\begin{equation}
 \overline{\cal A}_{x} = - {\frac{x}{2Ma^2}} \{ \vert E_0
  \vert^2{\,\rm Re}\,\alpha_{\bot} + \vert{\bf{E}}_0 {\bf{n}}
   \vert^2{\,\rm Re\,}(\alpha_{\parallel} - \alpha_{\bot} ) \},
    \label{eq24}
     \end{equation}
\begin{equation}
 \overline{\cal A}_{z}  = {\frac{k}{2M}}\{\vert E_{0}\vert^2
  {\,\rm Im}\,\alpha_{\bot} + \vert{\bf{E}}_0 {\bf{n}}
   \vert^2 {\,\rm Im\,}(\alpha_{\parallel} - \alpha_{\bot} ) \},
    \label{eq25}
     \end{equation}
where ${\bf E}^2_0$ is the density of electromagnetic field.
For the linear field polarization along $y$-axis, one must
change in Eq.~(\ref{eq24}) $x$ by $y$. Component labeled by $z$
is along with the spheroid rotation axis and another one
labeled by $x$ (or $y$) is transverse to this axis.

The real and imaginary parts
of the polarizability tensor can be written as \cite{G2}
\begin{widetext}
 \begin{equation}
   {\rm Re\,}\alpha_{\|\choose\bot} = \frac{V}{4\pi L_{\|\choose\bot}}
    \frac{\left((1-\xi_m)\omega^2-\omega^2_{\|\choose\bot}\right)
     \left(\omega^2- \omega^2_{\|\choose\bot}\right)+
      \left(2\omega\gamma_{\|\choose\bot}\right)^2 }{\left(\omega^2-
       \omega^2_{\|\choose\bot}\right)^2 + \left(2\omega
        \gamma_{\|\choose\bot}\right)^2},
         \label{eq26}
          \end{equation}
           \end{widetext}
and
\begin{equation}
 {\rm Im}\,\alpha_{\|\choose\bot}=\left(\frac{V}{4\pi L_{\|\choose\bot}}\right)
   \frac{2\omega^3\xi_m\gamma_{\|\choose\bot} }{\left(\omega^2-
    \omega^2_{\|\choose\bot}\right)^2+\left(2\omega\gamma_{\|\choose\bot}\right)^2},
     \label{eq27}
      \end{equation}
where we have introduced the notations
$$
V=\frac{4}{3}\pi R_{\|} R^2_{\bot},
$$
\begin{equation}
  \xi_m = \frac{\epsilon_m}{\epsilon_m+L_{\|\choose
   \bot}-L_{\|\choose\bot}\epsilon_m},
    \label{eq28}
     \end{equation}
\begin{equation}
 \omega^2_{\|\choose\bot} = \frac{L_{\|\choose\bot}}{
  \epsilon_m+L_{\|\choose\bot}-L_{\|\choose\bot}\epsilon_m} \;\omega^2_{pl},
   \label{eq29}
    \end{equation}
and
\begin{equation}
  \gamma_{\|\choose\bot}\equiv\gamma_{\|\choose\bot}(\omega) =
   \frac{2\pi L_{\|\choose\bot}}{\epsilon_m+L_{\|\choose\bot}-
    L_{\|\choose\bot}\epsilon_m} \sigma_{\|\choose\bot}(\omega)
     \label{eq30}
      \end{equation}
represents the half-width of the resonance curve for the light polarized
along ($\|$) or across ($\bot$) the rotation axis of the spheroid;
$\epsilon_m$ is the dielectric constant of the medium.

For the MN of a spherical form emersed in medium with
$\epsilon_m=1$, in the field of the same wave, one get
\begin{equation}
 \overline{\bf{\cal A}}_{\rm sph} = \frac{k}{2M} e^{-x^2/a^2}
  {E}^2_0\,{\rm Im}\,\alpha_{\rm sph} ,
   \label{eq31}
   \end{equation}
\begin{equation}
 \overline{\bf{\cal A}}_{\rm sph} = -\frac{x_i}{2Ma^2} e^{-x^2_i/a^2}
  {E}^2_0 \,{\rm Re}\,\alpha_{\rm sph},
   \label{eq32}
    \end{equation}
where $x_i = x, y$ and
\begin{equation}
 {\rm Re}\,\alpha_{\rm sph} = R^3\frac{(\varepsilon'-1)
  (\varepsilon'+2)+(4\pi\sigma/\omega)^2}
   {(\varepsilon'+2)^2+(4\pi\sigma/\omega)^2},
    \label{eq33}
     \end{equation}
\begin{equation}
 {\rm Im}\,\alpha_{\rm sph} = R^3\frac{12\pi\sigma/\omega}
   {(\varepsilon'+2)^2+(4\pi\sigma/\omega)^2},
    \label{eq34}
     \end{equation}
\begin{equation}
 \sigma = \frac{3}{16\pi}\frac{\upsilon_F}{R}
  \left(\frac{\omega_{\rm pl}}{\omega}\right)^2.
   \label{eq35}
    \end{equation}
 Here $R$ is the radius of a spherical MN, $\sigma$
 is its the high-frequency optical conductivity, and
 we take into account Eq.~(\ref{eq10}). Expressions
 (\ref{eq26}) and (\ref{eq27}), obtained for spheroidal
 MNs, clearly transforms into the corresponding
 expressions (\ref{eq33}) and (\ref{eq34}) for
 spherical MNs with the account for the equality
 $L_{\|}=L_{\bot}=1/3$. Then the conductivity becomes
 a scalar quantity, specified in the form (\ref{eq35}).
 At the plasma frequency the real part of
 the permittivity goes to zero.

Consider now the elliptical polarized Gaussian beam:
\begin{equation}
 {\bf{E}}_0 = ( {\bf{b}}_1 + i{\bf{b}}_2 ) e^{-(x^2+y^2)/2a^2} e^{ikz},
  \label{eq36}
   \end{equation}
\begin{equation}
   {\bf{b}}_1 = (b_{1},0,0), \quad {\bf{b}}_2 = (0,b_2,0).
    \label{eq37}
\end{equation}
In this case, after substituting Eqs.~(\ref{eq36}), (\ref{eq37}),
and (\ref{eq22}) into Eq.~(\ref{eq7}), we obtain
\begin{eqnarray}
 \overline{\cal A}_{i} &=& -{\frac{x}{2Ma^2}}e^{-(x^2+y^2)/a^2}
  \left\{(b_1^2 + b_2^2){\,\rm Re\,}\alpha_{\bot}
   \right.\nonumber\\& + &\left.
    [({\bf{n}} \,{\bf{b}}_1)^2 + ({\bf{n}} \, {\bf{b}}_2)^2]
     {\,\rm Re\,}(\alpha_{\parallel} - \alpha_{\bot} ) \right\},
      \label{eq38}
\end{eqnarray}
with $i=x,y$,
 \begin{eqnarray}
  \overline{\cal A}_{z} &=& \frac{k}{2M}e^{ - (x^2 + y^2)/a^2}
   \left\{(b_{1}^2 + b_2^2){\,\rm Im\,}\alpha_{ \bot}
    \right.\nonumber\\& + &\left.
     [({\bf{n}} \, {\bf{b}}_1)^2 + ({\bf{n}}\,
      {\bf{b}}_2)^2]{\,\rm Im\,}(\alpha_{\parallel} -
      \alpha_{ \bot}) \right\}.
       \label{eq39}
\end{eqnarray}
One should bear in mind that $\bf{n}$ is a unit vector directed along
the rotation axis of the ellipsoid. We see that in this case, the
acceleration depends on two angles -- between vectors $\bf{n}$
and ${\bf{b}}_1$ and between $\bf{n}$ and ${\bf{b}}_2$ ones.

If the components of the unit vector, appearing in Eq.~(\ref{eq22}),
in a spherical coordinate system are represented as
\begin{equation}
 n_x = \sin\theta \cos\varphi,\quad n_y = \sin\theta
  \sin\varphi\quad n_z = \cos\theta,
   \label{eq40}
    \end{equation}
then the products ${\bf nb}_1$ and ${\bf nb}_2$
in Eqs.~(\ref{eq38}) and (\ref{eq39}), respectively, become
\begin{equation}
 {\bf nb}_1 = {b}_1\sin\theta \cos\varphi,
  \quad {\bf nb}_2 = {b}_2\sin\theta \sin\varphi.
   \label{eq41}
    \end{equation}
In this case the ratio of the average values of the
MN acceleration, having spheroidal and spherical forms
in the direction of incidence of the radiation can be written,
using Eqs.~(\ref{eq39}) and (\ref{eq31}), as follows
\begin{eqnarray}
 &&\frac{\overline{\cal A}_z}{\overline{\cal A}_z|_{\rm sph}}=
  \frac{\rm Im\,\alpha_{\bot}}{\rm Im\,\alpha_{\rm sph}} \nonumber
   \\ && + \sin^2\theta\frac{({b}_1\cos\varphi)^2 +
    ({b}_2\sin\varphi)^2}{{b}_1^2+{b}_2^2}
     \frac{{\rm Im\,(\alpha_{\|}-
      \alpha_{\bot})}}{{\rm Im\,\alpha_{\rm sph}}}.
       \label{eq42}
        \end{eqnarray}
From Eq.~(\ref{eq42}), one can see that, contrary to the
MNs with a spherical form (when $\alpha_{\perp}=\alpha_{\parallel}$),
the acceleration components for the MNs with the ellipsoid-of-revolution
geometry acquire the dependence on the angle between the field direction
and the revolution axis of an spheroid. In addition, these acceleration components
depend on the MN's shape itself, which is determined by the depolarization
factors $L_j$ included in the diagonal components of the tensor $\alpha_{jj}$.
An analogous relation one can obtained as well for the ratio of the conservative
acceleration components $A_i$ if the imaginary parts of $\alpha$
in Eq.~(\ref{eq42}) will be replaced by they real parts.

In the case of circular polarization,
${\bf{b}}_1 = {\bf{b}}_2 = {\bf{b}}$, so that
\begin{equation}
 ({\bf{n}}\, {\bf{b}}_1)^2+({\bf{n}}\, {\bf{b}}_2)^2 =
  (n_x^2+n_y^2)b^2 = (1-n_z^2)b^2.
   \label{eq43}
\end{equation}
That is, in this case, only the dependence on the angle between
the vector $\bf{n}$\ and the direction of beam propagation survives.

Thus, similarly to the cases of plane-polarized and circularly
polarized light beams, the time-averaged acceleration of
non-spherical MN becomes angle-dependent.
In addition, this acceleration depends on the MN's shape
through the components $\alpha_{\bot}$ and $\alpha_{\|}$
of the polarization tensor; this dependence manifests itself
to the maximal extent in the infra-red range of the spectrum
(in the vicinity of the CO$_{2}$-laser frequency). For example,
taking $\omega_{pl}\approx 8 \times 10^{15}~{s}^{-1} $  for
gold and $\omega = 2\times 10^{14}~{s}^{-1}$ for the CO$_2$-laser
frequency, we obtain ${\varepsilon}^{\prime}\approx -1600$.
Therefore, the products $L_{\bot,\parallel}(\varepsilon_{\bot,
\|}-1)$ that enter into the denominators of formula
(\ref{eq16}) or (\ref{eq17}) are approximately equal
\begin{equation}
 L_{\parallel,\bot}(\varepsilon_{\parallel,\bot}-1)
  \approx -1600\,L_{\parallel,\bot}.
   \label{eq44}
    \end{equation}
Since the quantities $L_{\parallel}$ and $L_{\bot}$ may vary from 0 to 1
(provided that $2L_{\bot}+L_{\parallel}=1$), it is clear to which extent
quantity (\ref{eq24}) and, respectively, the quantities $\alpha_{\perp}$
and $\alpha_{\parallel}$ can be sensitive to the form of MN within this
range of frequencies.

\section{Acceleration due to Plasmon Resonance}

From Eqs.~(\ref{eq26}) and (\ref{eq27}) it is easy to determine
the real and imaginary parts of $\alpha_{\parallel,\bot}$ at
resonance frequencies
\begin{equation}
 {\rm Re\,}\alpha_{\parallel, \bot} = \frac{V}{4\pi L_{\parallel,\bot}}, \quad
  {\rm Im\,}\alpha_{\parallel, \bot} = \frac{V}{(4\pi L_{\parallel,\bot})^2}
   \frac{\omega_{\parallel, \bot}}{\sigma_{\parallel, \bot}}.
    \label{eq45}
     \end{equation}
For a predetermined frequency one can always chouse a geometric
form of the MN such that it will experience a resonant
increase of the acceleration with light pressure. In particular,
for MNs of spheroidal shape there exist two form of
the spheroid which can resonantly absorb radiation.
The opposite statement is also true: MN of an arbitrary
geometric shape will absorb resonantly at least one
frequency. The higher is the degree of symmetry of the MN,
the smaller is the number of resonant frequencies that it can
absorb. For example, a spherical MN has one, a spheroidal MN
-- two, and an ellipsoidal MN -- three resonant frequencies.

As one can see from Eq.~(\ref{eq42}), the value of the MN
acceleration ratio depends on both the angle of light incidence
$\theta$ and the light polarization angle $\varphi$.
Studies have shown that this ratio reaches a maximum value
at the angle of incidence equal to $\theta =\pi/2$.
Setting in Eq.~(\ref{eq42}) the angle $\theta$ equal to $\pi/2$,
let us investigate here how this relation changes for different
polarizations of the incident Gaussian beam with a change in
the shape of the MN. As an example, select the Cu
nanoparticle.

Fig.~1 illustrate the ratio of the acceleration of
spheroidal Cu nanoparticle to the acceleration of
spherical Cu nanoparticle under light pressure acting in
the direction of incidence of the laser beam, as a
function of the deviation of the shape of the MN
from spherical one. The frequency of laser beam was
chosen as $\omega = 2.9\times$10$^{15}$~s$^{-1}$, which
is close to the plasmon modes in copper and $\epsilon_m=1$.
As one can see, the resonant acceleration at that frequency
is experienced by MNs close to the spherical shape.

Fig.~2 shows the same dependence for the laterally directed
pressures. Here and below the calculation of acceleration components
are done using Eq.~(\ref{eq42}), as well as with using the analogous
expression obtained with the replacement ${\rm Im}\rightarrow{\rm Re}$
for the $i$-th pressure.

As is seen in Fig.~1, the acceleration under
laser radiation pressure on the Cu nanoparticle in the direction
of incidence of the laser beam at the plasmon resonance can be
hundreds of times greater than the pressure
experienced by a spherical Cu nanoparticle of equal volume.

In the lateral directions (Fig.~2) this excess is much less,
not exceeding of tens of times. In this case, the accelerations
ratio can be both positive or a negative value that speaks
for the attractive or repulsive nature of the pressure (the
"radiation wind") acting on MN in these direction.
It reaches the maximum in absolute value at angle $\varphi=\pi/2$.

\includegraphics[width=8.6cm]{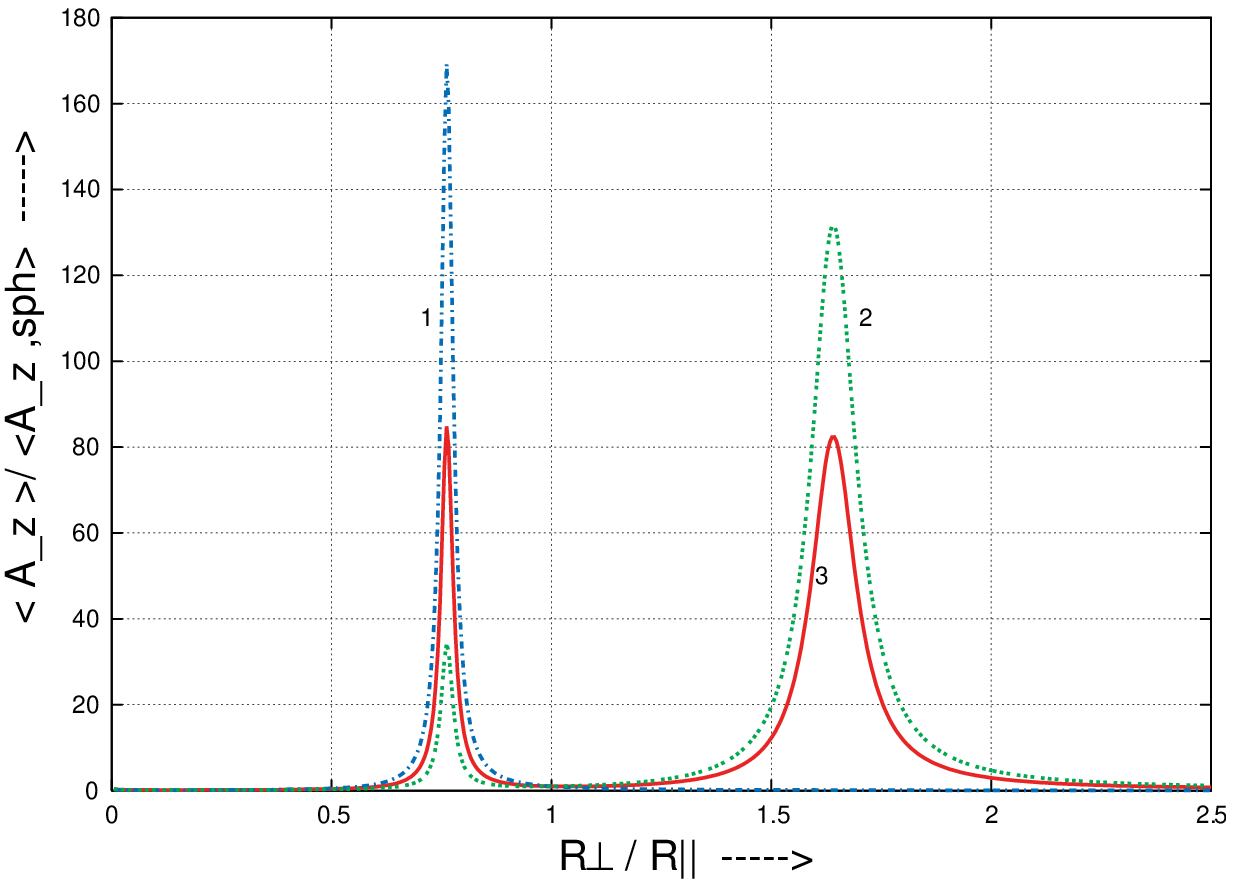}

\vskip-1mm\noindent{\footnotesize FIG.~1. {(Color online) The ratio of the acceleration
of spheroidal Cu nanoparticle to acceleration of spherical Cu nanoparticle
of equal volume with a radius of $100\AA$, as a function of
the Cu shape, in the direction of action of laser beam with
frequency $\omega\simeq 2.9\cdot 10^{15}~s^{-1}$ for different
polarization: curve~1 (short-dashed line) corresponds to the
linear polarization; curve~2 (long-dashed line) corresponds
to the elliptical polarization with $b_1/b_2=1/2$;
curve~3 (solid line) corresponds to the circular polarization.}}

\includegraphics[width=8.6cm]{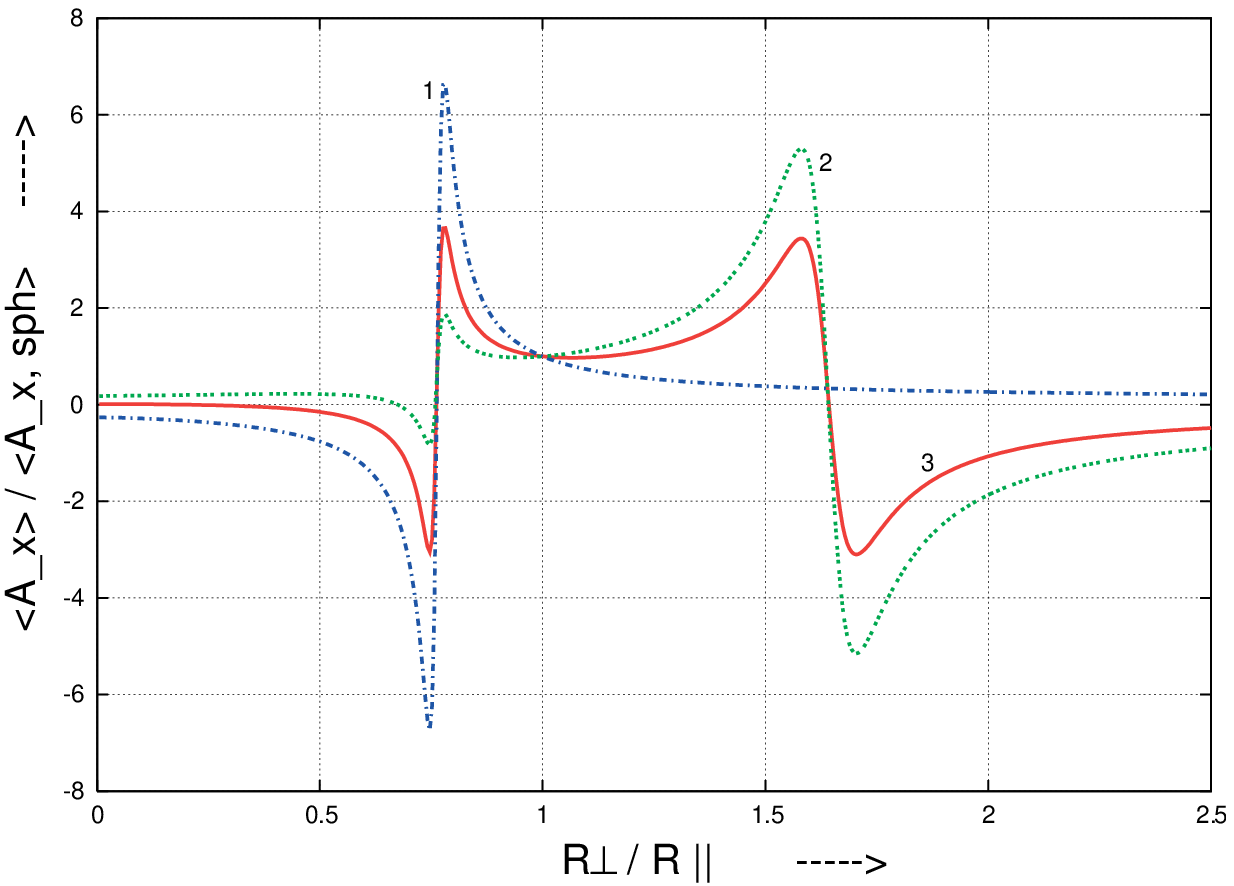}

\vskip-1mm\noindent{\footnotesize FIG.~2. {(Color online) The same 
as in Fig.~1 for the lateral $x$-direction of action of laser beam.}}

With increasing frequency of the incident radiation,
the plasmon resonance occurs for prolate MNs with
a greater ratio of $R_{\bot}/R_{\|}$, while for oblate
MNs -- with a smaller values of $R_{\bot}/R_{\|}$.
In the last case the pressure forces on the MN
fall off in absolute value.

It is also seen (Fig.~2) that in the lateral directions
together with the resonance for prolate MNs there is also
a resonance for oblate MNs. For low degrees of oblateness
this resonance is not suppressed by attenuation as it would
be in the case at CO$_2$-laser frequency. In the direction
of incidence of the radiation (along the z axis, Fig.~1), the
resonances appear in the form of peaks lying on each side
from the spherical shape $R_{\bot}/R_{\|}$=1.

At the pointed above plasmon frequency for Cu, the
resonance acceleration will be manifest itself
(in accordance with Eqs.~(\ref{eq29}) and (\ref{eq13}))
both for prolate and oblate MNs with the following
values of the ratio $R_{\bot}/R_{\|}=0.763$,
and $R_{\bot}/R_{\|}=1.64$, respectively.
Consequently, the left-hand peak (Fig.~1) and resonance
(Fig.~2) pertain to the prolate MN, while the
right-hand peak and resonance belong to the oblate MN.

\noindent\includegraphics[width=8.6cm]{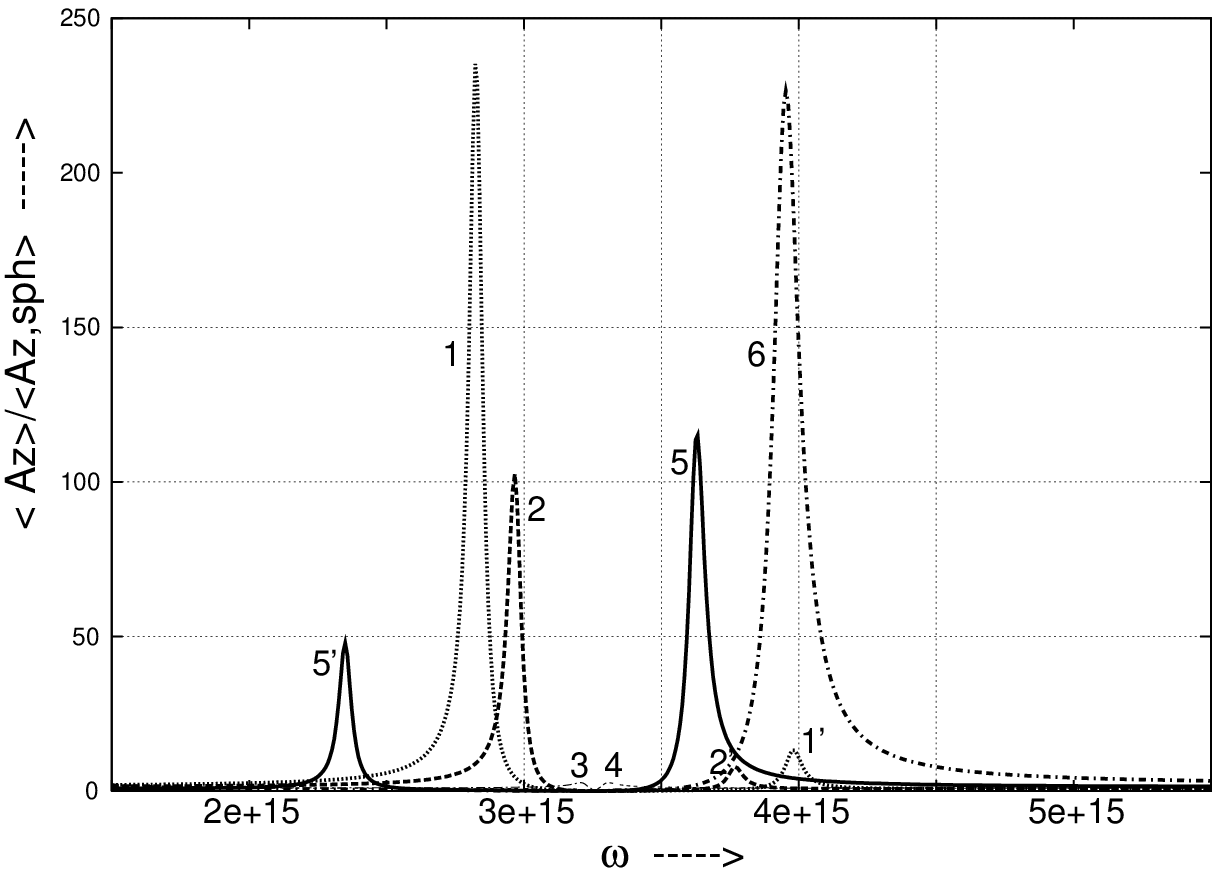}

\vskip-1mm\noindent{\footnotesize FIG.~3. {Frequency
dependence of the ratio of MN accelerations
in the direction of incidence laser beam for a spheroidal MN
{\it to} the acceleration of the spherical MN of equal volume
with the radius of $100\AA$, for nanoparticles with different
$R_{\bot}/R_{\|}$: 1.8(1), 1.5(2), 1.05(3), 0.95(4) 0.5(5) 0.1(6).
$\varphi=\pi/4$, $\theta=\pi/12$. $\epsilon_m=1.$}}%

As the angle $\theta$ decreases (with a
fixed angle $\varphi$), the resonance peaks
in the prolate MN are suppressed, whereas for
the oblate MN they reach maximum values.
For predetermined orientation (i.e., fixed $\varphi$
and $\theta$), enhancement or suppression of
the resonance peaks in MNs of different shapes
can be achieved by selecting the polarization
of the incident radiation\cite{KM}.

As the MN shape changes, there a shift of the plasmon
peaks of the acceleration occurs. To elucidate
the nature of these shifts with change in the degree
of oblateness or prolateness of the MN, in Fig.~3 we
have plotted the frequency dependence of the ratio of MN
accelerations with different shapes at fixed angles.
The weak peaks 3 and 4 in this figure pertain to the
MNs which are close to spherical in shape.

As can be seen from Fig.~3, the oblate MNs manifest
resonant acceleration at longer wavelengths (curves~1,2)
in comparison with a spherical MN, and
the prolate MNs -- at shorter wavelengths (curves~5,6).
Here the tails of the peaks of the oblate MNs extend
toward the long-wavelength side of the spectrum, while
those of the prolate MNs extend toward the short-wavelength
side. As the flatness of the MN increases (curve~1),
the particle acceleration peak increases in absolute value
and shifts to the longer wavelengths, while with increasing
in elongation of the MN (curve~6) there is,
in addition to increase its acceleration, the shift
of the peak to the shorter wavelengths of the spectrum.

The peaks labeled
by numbers with a prime in Fig.~3 arise together with the
peaks labeled with the corresponding unprimed numbers; as
can be seen from Eq.~(\ref{eq29}), this is due to their
falling into a given range of $R_{\bot}/R_{\|}$ values.
Their intensity depends on two factors: the orientation of
the MN with respect to the incident radiation and/or
polarization of the radiation. In the given example the
height of the primed peaks can be controlled by means
of the angle $\theta$; they disapper when $\theta\rightarrow 0$.

It should be noted that for prolate MNs, the growth
of intensity and the peak displacement with an increase in
the prolateness of the MN goes to saturation, after which further
increase in the prolateness leads to a drop in the intensity
of radiation pressure on the MN, and the peak displacement does
not occur. The evaluations show that already for $R_{\bot}/R_{\|}$=1/64,
the peak becomes rather wide, and its displacement is not very noticeable
in comparison with that for $R_{\bot}/R_{\|}$=1/32, for example.

\section{Conclusion}
We have obtained the simple analytical expressions for the
acceleration of the spheroidal MN which is exerted by a laser
beam averaged over a period of the incident wave. It is shown that
the acceleration components can substantially depend on the shape of
the MN as well as on the angles that define its orientation
both relative to the direction of the incident radiation and to
the polarization of the beam.

The behavior of the MN acceleration near plasmon resonances
in spheroidal MNs was also investigated, depending on the change
in the shape of MN and its orientation.
The shift of the resonance peak of the acceleration
to longer wavelengths of the spectrum is detected for more
oblate MNs and to shorter wavelengths -- for more prolate ones.
We have established that the value of the accelerations under
laser beam action on spheroidal MN can differ
by orders of magnitude from the accelerations of
the spherical MN of the same volume.

\begin{acknowledgments}
Author is grateful to the Program of the Fundamental Research of the
Department of Physics and Astronomy of the National Academy of Sciences
of Ukraine (NASU) (0120U100858) for financial support of this work.
\end{acknowledgments}

\end{document}